\documentclass[11pt]{revtex4-1}
\usepackage{bm}
\usepackage{color,graphicx}
\usepackage{tensor}
\usepackage{amsmath}
\begin{document}
\newcommand{\boldnabla}{\bm{\nabla}}
\newcommand{\boldtheta}{\bm{\theta}}
\newcommand{\angmom}{\bm{\mathcal{J}}}


\title{Glauber P-representations for fermions}

\author{Stephen M.~Barnett}

\address{School of Physics and Astronomy, University of Glasgow,
Glasgow G12 8QQ, United Kingdom}

\author{Bryan J.~Dalton}

\address{Centre for Quantum Science and Technology Theory, Swinburne University of Technology,
Melbourne, Victoria, 3122, Australia}


\date{\today}


\begin{abstract}
\noindent The Glauber-Sudarshan P-representation for bosons is well-known within quantum optics, and is widely applied to problems involving photon statistics.  Less
familiar, perhaps, is its fermionic counterpart, introduced by Cahill and Glauber.  We present a derivation of both the bosonic and fermionic distributions and, in doing so, demonstrate the reason for the existence of two distinct fermionic forms and the relationship between these.  We consider both single mode systems and also multiparticle
systems with many modes.  Expressions for the moments involving products of mode annihilation and creation operators are obtained.  For simplicity only one type of
boson or fermion will be considered, but generalising to more types is straightforward.

\end{abstract}


\maketitle



\section{Introduction}

The coherent states of light were first introduced in the context of the quantum theory of optical coherence
\cite{Glauber63,Sudarshan63,Glauber63a,Glauber63b,Louisell,Mehta}.  These states are remarkable for representing
the closest approximation, within quantum theory, to fully coherent classical light.  They form, moreover, an
overcomplete set of states and this property has led to their widespread application in quantum optics. 
Of particular interest to us is the representation of the state of a quantised field mode of the form 
\cite{footnote}
\begin{equation}
\label{Eq1}
\hat{\rho} = \int d^2\alpha P(\alpha,\alpha^*)|\alpha\rangle\langle\alpha| \, ,
\end{equation}
where $|\alpha\rangle$ is the coherent state parametrised by the complex number $\alpha$ and the 
real function $P(\alpha)$ is a real quasi-probability distribution, and is a function of both $\alpha$
and $\alpha^*$ \cite{Radmore}.  This P-representation of the density operator has been widely 
employed in quantum optics, particularly in the evolution of open systems,  and is usually introduced 
as an ansatz in the form of Eq. (\ref{Eq1})
\cite{Radmore,Klauder,SSL,Pierre,Gardiner,Carmichael,Dan,MW,Zubairy}.  It is possible, however, to derive
this using a theorem due to Weyl \cite{Weyl} and presented in the context of quantum optics
by Cahill and Glauber \cite{CG69}.  This derivation seems not to have not been widely appreciated
and does not appear in most textbooks on quantum optics.  We start, therefore, with a presentation
of this derivation.  Note, however, that not every density operator possess a P representation
\cite{CG69a,Cahill69,Cahill69a}.
Before proceeding, we point out that there a number of ways of introducing 
phase space distributions, sometimes based on different choices of coherent states.  These include 
working with coherent states based on particle-hole pairs \cite{Rowe}, group theoretical methods \cite{Brif}
and spin or atomic coherent states \cite{Agarwal}.  We note also, that a fermionic Wigner function 
has been introduced \cite{AbeSuziki} as has a phase space description for coupled boson-fermion
systems using supercoherent states \cite{Abe}.  A comprehensive introduction to a wide variety of
coherent states can be found in \cite{Perelomov}.

The coherent states for fermions are less familiar, certainly to those in the field of quantum optics,
but are play an important role in the many-body physics and quantum field theory of fermions.  In
these states the complex c-numbers $\alpha$ for bosonic coherent states are replaced by anticommuting
Grassmann variables \cite{Berezin,Rivers,Itzykson,Zinn-Justin,Kleinert,Mandl,Bailin,Aitchison,DJB}.  Remarkably, however,
two quite distinct forms have been reported for the corresponding P-representation of a fermionic
state.  When simplified to a single fermionic mode, these represent the density operator either as
\begin{equation}
\label{Eq2}
\hat{\rho} = \int d^2g \: \phi(g,g^*)|g\rangle\langle g|
\end{equation}
or as
\begin{equation}
\label{Eq3}
\hat{\rho} = \int d^2g \: P(g,g^*)|g\rangle\langle -g| \, ,
\end{equation}
where $g$ and $g^*$ are Grassmann variables.  The first of these is essentially that employed by Plimak {\it et al} in a 
study of Cooper-like pairing of fermions \cite{Plimak}, and the latter was introduced by Cahill and 
Glauber \cite{CG99}.  We derive both of these and, in doing so, uncover the reason
for the existence of two such representations, when only a single P representation appears for bosons.
The two P functions, appearing in the above representations of the density operator, are quite distinct 
and this is the reason for writing the first as $\phi(g,g^*)$.  



\section{P-representation for bosons}

The P-representation of a density operator for a single boson mode is given in Eq. (\ref{Eq1}).  Our task is to derive
this expression and, in doing so, to arrive at the correct form of the function $P(\alpha,\alpha^*)$.  Note that we do not take the
more familiar path of supposing this relation to be true and deriving the requisite properties of $P(\alpha,\alpha^*)$ from this
assumption.  We follow closely the method of Cahill and Glauber \cite{CG69} in deriving, first a completeness relation for
the displacement operators, and then using this to derive Eq. (\ref{Eq1}).  We start with just a single mode, but generalise
to multi-mode states at the end of the section.  For simplicity we consider only one type of boson.

\subsection{Completeness of the displacement operators}

The Glauber displacement operator has the form \cite{Radmore}
\begin{equation}
\label{Eq4}
\hat{D}(\alpha,\alpha^*) = \exp(\alpha\hat{a}^\dagger - \alpha^*\hat{a})  \, ,
\end{equation}
which can be written in the normally or antinormally ordered forms:
\begin{equation}
\label{Eq4a}
\hat{D}(\alpha,\alpha^*) = \exp(\alpha\hat{a}^\dagger)\exp(-\alpha^*\hat{a})\exp\left(-\frac{|\alpha|^2}{2}\right)
= \exp(-\alpha^*\hat{a})\exp(\alpha\hat{a}^\dagger)\exp\left(\frac{|\alpha|^2}{2}\right) \, .
\end{equation}
This unitary operator acts to generate the coherent state $|\alpha,\alpha^*\rangle$ from the vacuum (or zero particle) state:
\begin{equation}
\label{Eq5}
|\alpha,\alpha^*\rangle = \hat{D}(\alpha,\alpha^*)|0\rangle \, .
\end{equation}
To condense the notation, we follow common practice and replace $|\alpha,\alpha^*\rangle$ and $\hat{D}(\alpha,\alpha^*)$
by $|\alpha\rangle$ and $\hat{D}(\alpha)$ respectively.
Our first task is to show that an operator $\hat{F}$ can be written in the form
\begin{equation}
\label{Eq6}
\hat{F} = \int \frac{d^2\xi}{\pi}f(\xi)\hat{D}(-\xi) \, ,
\end{equation}
where $d^2\xi = d\xi'd\xi''$ corresponds to integrating the real and imaginary parts of over the whole of the complex
$\xi$ plane.  Here the function $f(\xi)$ is related to the operator $\hat{F}$ by 
\begin{equation}
\label{Eq7}
f(\xi) = {\rm Tr}[\hat{F}\hat{D}(\xi)] \, .
\end{equation}
We can think of the expansion in Eq. (\ref{Eq6}) as the operator analogue of a Fourier expansion or transform,
in which the operator is expanded in terms of a complete set of displacement operators in place of a complete
set of functions of the form $e^{ikx}$ \cite{Weyl}.
The representation of the operator $\hat{F}$ in Eq (\ref{Eq6}) is possible when $f(\xi)$ is square-integrable or,
equivalently, when the operator $\hat{F}$ has finite Hilbert-Schmidt norm:
\begin{equation}
\label{Eq8}
\int\frac{d^2\xi}{\pi}|f(\xi)|^2 = {\rm Tr}(\hat{F}^\dagger\hat{F}) < \infty .
\end{equation}

To proceed with the proof, we follow Cahill and Glauber \cite{CG69} by introducing a function of four complex variables
\begin{equation}
\label{Eq9}
I(\alpha,\beta,\gamma,\delta) = \int\frac{d^2\xi}{\pi} \langle\beta|\hat{D}(\xi)|\alpha\rangle\langle\gamma|\hat{D}(-\xi)|\delta\rangle
\end{equation}
It is straightforward to use the normally ordered form of the displacement operator, Eq (\ref{Eq4a}), and the overlap of two coherent states \cite{Radmore} 
to write this function in the form
\begin{eqnarray}
\label{Eq10}
I(\alpha,\beta,\gamma,\delta) &=& \langle\beta|\alpha\rangle\langle\gamma|\delta\rangle\int\frac{d^2\xi}{\pi} e^{-|\xi|^2}e^{\xi\beta^*-\xi^*}e^{-\xi\gamma^*+\xi^*\delta} \, .
\nonumber \\
&=& \langle\gamma|\alpha\rangle\langle\beta|\delta\rangle \, .
\end{eqnarray}
As the coherent states are complete (or more precisely overcomplete \cite{Radmore,Cahill65}), 
we have an identity between two operators if their coherent
state matrix elements are identical.  It follows, therefore, that we can write the outer product of operator $|\alpha\rangle\langle\beta|$
in the form
\begin{equation}
\label{Eq11}
|\alpha\rangle\langle\beta| = \int\frac{d^2\xi}{\pi}\langle\beta|\hat{D}(\xi)|\alpha\rangle\hat{D}(-\xi) \, .
\end{equation}
Any bounded operator, $\hat{F}$, can be written in the form
\begin{eqnarray}
\label{Eq12}
\hat{F} &=& \int \frac{d^2\alpha d^2\beta}{\pi^2} |\alpha\rangle\langle\alpha|\hat{F}|\beta\rangle\langle\beta| \nonumber \\
&=& \int \frac{d^2\alpha d^2\beta}{\pi^2} |\alpha\rangle\langle\beta| \, \langle\alpha|\hat{F}|\beta\rangle \, ,
\end{eqnarray}
where we have used the resolution of the identity operator in terms of coherent states:
\begin{equation}
\label{Eq13}
\int\frac{d^2\alpha}{\pi}|\alpha\rangle\langle\alpha| = \hat{\rm I} \, .
\end{equation}
It then follows that
\begin{eqnarray}
\label{Eq14}
\hat{F} &=& \int\frac{d^2\alpha d^2\beta d^2\xi}{\pi^3}\langle\beta|\hat{D}(\xi)|\alpha\rangle\hat{D}(-\xi) \, \langle\alpha|\hat{F}|\beta\rangle
\nonumber \\
&=& \int\frac{d^2\alpha d^2\beta d^2\xi}{\pi^3}\langle\beta|\hat{D}(\xi)|\alpha\rangle\langle\alpha|\hat{F}|\beta\rangle \hat{D}(-\xi)
\nonumber \\
&=&  \int\frac{d^2\beta d^2\xi}{\pi^2}\langle\beta|\hat{D}(\xi)\hat{F}|\beta\rangle \hat{D}(-\xi) \, .
\end{eqnarray}
Finally, we note that the integral over $\beta$ corresponds to the trace and so we find
\begin{equation}
\label{Eq14a}
\hat{F} = \int\frac{d^2\xi}{\pi}{\rm Tr}[\hat{F}\hat{D}(\xi)]\hat{D}(-\xi) \,  ,
\end{equation}
which is the expression in Eq. (\ref{Eq6}).  For fermions we shall find and make use of a similar theorem, but in that case
the final step leading to a trace operation does not hold.

\subsection{Expressing the density operator in terms of the P-function}

We can complete our derivation of the P-representation of the density operator by applying the general result in Eq. (\ref{Eq6}) to
the density operator.  This is allowable because the density operator will always have a finite Hilbert-Schmidt norm: 
${\rm Tr}(\rho^2) \leq 1$.  Direct application of our theorem gives the form
\begin{equation}
\label{Eq15}
\hat{\rho} = \int\frac{d^2\xi d^2\alpha}{\pi^2}{\rm Tr}[\hat{\rho}\hat{D}(\xi)]\hat{D}(-\xi) \, .
\end{equation}
To obtain the P-representation form from this we first write the displacement operators in an ordered form, normal order for the first and
antinormal order for the second:
\begin{equation}
\label{Eq16}
\hat{\rho} = \int\frac{d^2\xi}{\pi}{\rm Tr}[\hat{\rho}e^{\xi\hat{a}^\dagger}e^{-\xi^*\hat{a}}]e^{\xi^*\hat{a}}e^{-\xi\hat{a}^\dagger} \, .
\end{equation}
All that remains is to insert the identity in the form given in Eq (\ref{Eq13}) between the final two operators:
\begin{eqnarray}
\label{Eq17}
\hat{\rho} &=& \int\frac{d^2\xi}{\pi}{\rm Tr}[\hat{\rho}e^{\xi\hat{a}^\dagger}e^{-\xi^*\hat{a}}]e^{\xi^*\hat{a}}
\int\frac{d^2\alpha}{\pi}|\alpha\rangle\langle\alpha|e^{-\xi\hat{a}^\dagger} \nonumber \\
&=& \int d^2\alpha P(\alpha,\alpha^*) |\alpha\rangle\langle\alpha| \, ,
\end{eqnarray}
where the P function is
\begin{equation}
\label{Eq18}
P(\alpha,\alpha^*) = \int\frac{d^2\xi}{\pi^2}{\rm Tr}[\hat{\rho}e^{\xi\hat{a}^\dagger}e^{-\xi^*\hat{a}}]e^{\alpha\xi^*-\alpha^*\xi} \, .
\end{equation}
Note that this function is the Fourier transform of the normally ordered characteristic function \cite{Radmore,CG69a}
\begin{equation}
\label{Eq19}
\chi(\xi,\xi^*) = {\rm Tr}[\hat{\rho}e^{\xi\hat{a}^\dagger}e^{-\xi^*\hat{a}}]
\end{equation}
and the P-function will be well-behaved only if this Fourier transform exists.  In the study of non-classical states of light
this feature has often been used to distinguish between classical and intrinsically quantum properties of light \cite{MW}.
We note that the expectation values of normally ordered functions of the creation and annihilation operators 
are readily expressed in terms of the P-function.  In general we have
\begin{equation}
\label{Eq20}
\langle\hat{a}^{\dagger n}\hat{a}^m\rangle = \int d^2\alpha P(\alpha,\alpha^*) \alpha^{*n}\alpha^m \, .
\end{equation}
We note, in particular, that this expression includes the normalisation of the P-function (for $n=0=m$):
\begin{equation}
\label{Eq20aa}
\int d^2\alpha P(\alpha,\alpha^*) = 1\, .
\end{equation}
We shall encounter similar expressions for our fermionic P representations.

It is often the case that we need to describe multimode field states and for this purpose we need a more general form 
for our P-representation.  To this end we introduce a complex number, $\alpha_i$, for each of the modes.  The completeness
relation for the boson coherent states is
\begin{equation}
\label{Eq20a}
\int \frac{d^2\mbox{\boldmath$\alpha$}}{\pi^n} |\mbox{\boldmath$\alpha$}\rangle\langle\mbox{\boldmath$\alpha$}| = \hat{\rm I} \, ,
\end{equation}
where {\boldmath$\alpha$} is a shorthand for $\alpha_1, \alpha_2, \cdots \alpha_n$, so that $|\mbox{\boldmath$\alpha$}\rangle$
is the multimode state $|\alpha_1\rangle|\alpha_2\rangle\cdots|\alpha_n\rangle$ and $d^2\mbox{\boldmath$\alpha$} =
\prod_i d^2\alpha_i$.  The multimode P-function will then be a function of all of the c-numbers $\alpha_i$ and $\alpha^*_i$.  
The required form for an $n$ mode state is simply
\begin{equation}
\label{Eq21}
\rho = \int d^2\mbox{\boldmath$\alpha$} P(\mbox{\boldmath$\alpha$},\mbox{\boldmath$\alpha$}^*)|\mbox{\boldmath$\alpha$}\rangle\langle\mbox{\boldmath$\alpha$}| \, .
\end{equation}
The form of this multimode P-function is the natural analogue of the single-mode form given in Eq (\ref{Eq18}):
\begin{equation}
\label{Eq21a}
P(\mbox{\boldmath$\alpha$},\mbox{\boldmath$\alpha$}^*) = \int\prod_i\left(\frac{d^2\xi_i}{\pi^2}\right){\rm Tr}\left[\hat{\rho}
\exp(\mbox{\boldmath$\xi$}\cdot\hat{\bf a}^\dagger)\exp(-\mbox{\boldmath$\xi$}^*\hat{\bf a})\right]
\exp(\mbox{\boldmath$\alpha$}\cdot\mbox{\boldmath$\xi$}^* - \mbox{\boldmath$\alpha$}^*\cdot\mbox{\boldmath$\xi$}) \, ,
\end{equation}
where $\mbox{\boldmath$\xi$}\cdot\hat{\bf a}^\dagger = \sum_i\xi_i\hat{a}_i^\dagger$, with similar expressions for the other
terms.  It is straightforward to verify that the multimode P-function is normalised:
$\int d^2\mbox{\boldmath$\alpha$} P(\mbox{\boldmath$\alpha$},\mbox{\boldmath$\alpha$}^*) = 1$.  
The normally ordered moments are given by the
integral of $P(\mbox{\boldmath$\alpha$},\mbox{\boldmath$\alpha$}^*) $ multiplied by a product of the $\alpha_i$ and $\alpha_j^*$, with these 
variables replacing the annihilation and creation operators $\hat{a}_i$ and $\hat{a}^\dagger_j$ respectively.


\section{P-representations for fermions}

The title of this section deliberately employs the plural `representations' as the construction we have employed above
gives rise, in the fermionic case, to more than one expression for any given density operator.  We shall find that the 
origin of this non-uniqueness lies in the existence of more than one integral representation of the identity in terms of 
fermionic coherent states.  We follow closely the approach outlined in the preceding section for bosons, starting with
a completeness relation for fermionic coherent states and following this with a derivation of the P-representations.
As with the bosonic case, we begin with just a single mode before generalising to multi-mode states at the end of the section.
Again, for simplicity, we consider only systems with one type of fermion.

\subsection{Grassmann variables and fermionic coherent states}

Before proceeding with our derivation of the P-representation, we present a brief review of the principal properties of
Grassmann variables and of the coherent states constructed with them.  Fuller accounts can be found in 
\cite{Berezin,DJB,Plimak,CG99,Bowden,Junker,Combescure}.  Grassmann variables are, in effect, anti-commuting `numbers' 
\footnote{Grassmann variables were introduced by Hermann Grassmann in his ground-breaking work on linear
algebra \cite{GrassmannBook}.  An account
of this work and its importance may be found in \cite{Fearnley-Sander}.}.  
If $g$ and $h$ are any two Grassmann variables then 
\begin{equation}
\label{Eq21b}
gh = -hg \, ,
\end{equation}
which implies that the square of any Grassmann variable is zero.  It follows that expressions containing an even or an odd
number of Grassmann variables behave rather differently: an even expression commutes with other Grassmann variables,
but an odd one anti-commutes: $g_1g_2g_3 = g_2g_3g_1$, but $g_1g_2g_3g_4 = -g_2g_3g_4g_1$.  

We can define complex conjugation for Grassmann variables by introducing $g^*$ as the complex conjugate of $g$.  It is
convenient to define this operation to reverse the order of the variables:
\begin{equation}
(gh)^* = h^*g^* \, ,
\end{equation}
which is reminiscent of the Hermitian conjugation operation for matrices and operators. We can define differentiation
and integration for Grassmann variables \cite{Berezin,DJB}.  We shall require only Grassmann integration, for which the rules
are
\begin{eqnarray}
\label{Eq22}
\int dg &=& 0 \nonumber \\
\int dg \: g &=& 1 \, . 
\end{eqnarray}
When there is more than one Grassmann variable involved we need to be careful with the order both of the
variables in the integrand and also with the differentials:
\begin{eqnarray}
\label{Eq23}
\int dhdg \: gh &=& 1 \nonumber \\
\int dhdg \: hg &=& \int dh dg (-gh) = -1 \ .
\end{eqnarray}

We denote the fermionic annihilation and creation operators by $\hat{c}$ and $\hat{c}^\dagger$, which satisfy the
anticommutation relation
\begin{equation}
\label{Eq23a}
\{\hat{c},\hat{c}^\dagger\} = \hat{c}\hat{c}^\dagger + \hat{c}^\dagger\hat{c} = 1 \, :
\end{equation}
The two possible fermionic number states are the vacuum, or no particle, state $|0\rangle$ and the one particle state
$|1\rangle$:
\begin{eqnarray}
|1\rangle &=& \hat{c}^\dagger |0\rangle \nonumber \\
|0\rangle &=& \hat{c}|1\rangle \, .
\end{eqnarray}
The properties of the annihilation and creation operators are complicated by the fact that they anticommute with the Grassmann 
variables in that
\begin{equation}
g\hat{c} = -\hat{c}g  \qquad g\hat{c}^\dagger = -\hat{c}^\dagger g \, .
\end{equation}
These features mean that we need to be careful with the ordering of both our Grassmann variables and our operators.

The fermionic coherent states are characterised by a Grassmann variable $g$ rather than a c-number as in the
bosonic case.  Specifically, they are generated from the vacuum state by means of a unitary transformation 
generated by a fermionic displacement operator:
\begin{equation}
\label{Eq24}
\hat{D}(g,g^*) = \exp(\hat{c}^\dagger g - g^*\hat{c}) \, ,
\end{equation} 
which can be expressed in normal or antinormal ordered forms as
\begin{equation}
\label{Eq25}
\hat{D}(g,g^*) = \exp(\hat{c}^\dagger g)\exp( - g^*\hat{c}) \exp\left(-\frac{g^*g}{2}\right) = \exp( - g^*\hat{c})\exp(\hat{c}^\dagger g)\exp\left(\frac{g^*g}{2}\right)  \, .
\end{equation} 
We note that these have same form as their bosonic counterparts.  As with their bosonic counterparts, Eq (\ref{Eq5}),
the fermionic coherent states are given by 
\begin{equation}
|g,g^*\rangle = \hat{D}(g,g^*)|0\rangle \, .
\end{equation}
Again we simplify the notation by using $|g\rangle$ and $\hat{D}(g)$.
It follows that our fermionic coherent states have the form
\begin{equation}
\label{Eq26}
|g\rangle = \left(1 - \frac{g^*g}{2}\right)(|0\rangle - g|1\rangle) =  \left(1 - \frac{g^*g}{2}\right)|0\rangle - g|1\rangle \, .
\end{equation}
The coherent states are right-eigenstates of the annihilation operator with eigenvalue $g$:
\begin{equation}
\label{Eq27}
\hat{c}|g\rangle = -\hat{c} g \hat{c}^\dagger |0\rangle = g\hat{c}\hat{c}^\dagger|0\rangle = g|0\rangle = g|g\rangle 
\end{equation}
and are left-eigenstates of the creation operator with eigenvalue $g^*$: 
\begin{equation}
\label{Eq27a}
\langle g|\hat{c}^\dagger = \langle g|g^* \neq g^*\langle g| \, .
\end{equation}
The overlap between two fermionic coherent states is 
\begin{equation}
\label{Eq28}
\langle g|h\rangle = \left(1 - \frac{g^*g}{2}\right)\left(1 - \frac{h^*h}{2}\right)(1 + g^*h) \, .
\end{equation}
We can write this in terms of exponential functions 
\begin{equation}
\label{Eq29}
\langle g|h\rangle = \exp\left(- \frac{g^*g}{2}\right)\exp\left(- \frac{h^*h}{2}\right)\exp(g^*h) \, ,
\end{equation}
which has the same form as for the bosonic coherent states \cite{Radmore}.  We may require that the vacuum state is an 
even Grassmann function so that $h|0\rangle = |0\rangle h$.  It follows that the one particle state, $|1\rangle = \hat{c}^\dagger|0\rangle$, 
is odd, and it follows that our coherent states are even Grassmann functions and therefore commute with Grassmann variables:
\begin{equation}
\label{Eq29a}
h|g\rangle = |g\rangle h  \qquad h\langle g| = \langle g|h \, .
\end{equation}

Finally, we can resolve the identity in terms an integral over the coherent states:
\begin{eqnarray}
\label{Eq30}
\int dg^*dg \: |g\rangle\langle g| &=& \int dg^*dg (1 - g^*g)(|0\rangle - g|1\rangle)(\langle 0| - \langle 1|g^*)
\nonumber \\
&=&  \int dg^*dg (1 + gg^*)[|0\rangle\langle 0| + g|1\rangle\langle 1|g^*] \nonumber \\
&=& |0\rangle\langle 0| + |1\rangle\langle 1| = \hat{\rm I} \, ,
\end{eqnarray}
which is the natural analogue of the bosonic expression given in Eq (\ref{Eq13}).  In contrast 
with the bosonic case, however, this is not the only way to resolve the identity in terms of the
coherent states.  In particular we can also express the identity as an integral over the operators
$|g\rangle\langle -g|$ in the form
\begin{eqnarray}
\label{Eq33a}
\int dg^*dg (2gg^*-1)|g\rangle\langle -g| &=& \int dg^*dg (2gg^* - 1)(1 + gg^*)(|0\rangle - g|1\rangle)
(\langle 0| + \langle 1|g^*) \nonumber \\
&=&  \int dg^*dg (gg^* - 1)(|0\rangle\langle 0| - g|1\rangle\langle 1|g^*) \nonumber \\
&=& |0\rangle\langle 0| + |1\rangle\langle 1| = \hat{\rm I} \, .
\end{eqnarray}
We shall find that it is this existence of a second resolution of the identity that leads to 
the existence of two P-representations for fermions \cite{footnote1}.

\subsection{Completeness of the displacement operators}

We follow the derivation of the completeness of the bosonic displacement operators, given above, by
introducing the Grassmann function
\begin{equation}
\label{Eq31}
I(g_1,g_2,g_3,g_4) = \int dh^*dh \: \langle g_2|\hat{D}(h)|g_1\rangle\langle g_3|\hat{D}(-h)|g_4\rangle \, .
\end{equation}
Careful evaluation of the matrix elements and Grassmann integrals gives the same result as found 
for bosons:
\begin{equation}
\label{Eq32}
I(g_1,g_2,g_3,g_4) = \langle g_2|g_4\rangle\langle g_3|g_1\rangle \, .
\end{equation}
Two operators are equivalent if their coherent state matrix elements are the same:
\begin{equation}
\label{Eq33}
\langle g|\hat{A}|h\rangle = \langle g|\hat{B}|h\rangle \quad \Rightarrow \quad \hat{A} = \hat{B} \, .
\end{equation}
It follows that 
\begin{equation}
\label{Eq34}
|g_1\rangle\langle g_2| = \int dh^*dh \: \langle g_2|\hat{D}(h)|g_1\rangle \hat{D}(-h) \, .
\end{equation}
To proceed we can insert the identity operator either before or after the displacement operator 
$\hat{D}(h)$ and use the fact that the overlap between two coherent states is an even Grassmann
function to write
\begin{eqnarray}
\label{Eq35}
|g_1\rangle\langle g_2| &=& \int dh^*dh \int dk^*dk \langle g_2|\hat{D}(h)|k\rangle\langle k| g_1\rangle 
\hat{D}(-h) \nonumber \\
&=& \int dh^*dh \int dk^*dk \langle k| g_1\rangle\langle g_2|\hat{D}(h)|k\rangle 
\hat{D}(-h) \, ,
\end{eqnarray}
where $k$ and its complex conjugate $k^*$ are a further pair of Grassmann variables.
We can extract a general relationship for the four independent single-mode operators,
$|0\rangle\langle 0|$, $|0\rangle\langle 1|$, $|1\rangle\langle 0|$ and $|1\rangle\langle 1|$,
and hence for any operator.  We can achieve this either by comparing the coefficients of
the Grassmann variables or, more formally, by Grassmann integration.  It follows that 
for a general single-mode operator, $\hat{F}$ we have the identity
\begin{equation}
\label{Eq36}
\hat{F} = \int dh^*dh \int dk^*dk \langle k|\hat{F}\hat{D}(h)|k\rangle\hat{D}(-h) \, ,
\end{equation}
which has the same form as that given in Eq (\ref{Eq14}) for a bosonic operator.  Note,
however, that unlike in the bosonic case, the $k$-integral does not, in general, correspond to the
trace of the operator product $\hat{F}\hat{D}(h)$.

\subsection{Expressing the density operator in terms of the P-functions}

To derive the P representations we follow our analysis for bosons by ordering the two
displacement operators, $\hat{D}(h)$ and $\hat{D}(-h)$ to give
\begin{equation}
\label{Eq37}
\hat{\rho} = \int dh^*dh \int dk^*dk \langle k| \hat{\rho} e^{\hat{c}^\dagger h}e^{-h^*\hat{c}}|k\rangle
e^{h^*\hat{c}}e^{-\hat{c}^\dagger h} \, .
\end{equation}
At this point we can insert the identity operator, expressed in terms of coherent states, between
the final two operators and hence obtain expressions for our two Grassmann functions
$\phi(g,g^*)$ and $P(g,g^*)$.  For $\phi(g,g^*)$ we insert the identity as given in Eq (\ref{Eq30}) so that
\begin{eqnarray}
\label{Eq48}
\hat{\rho} &=& \int dh^*dh \int dk^*dk \langle k| \hat{\rho} e^{\hat{c}^\dagger h}e^{-h^*\hat{c}}|k\rangle
e^{h^*\hat{c}}\int dg^*dg |g\rangle\langle g| e^{-\hat{c}^\dagger h}  \nonumber \\
&=& \int dg^*dg \int dh^*dh \int dk^*dk \langle k| \hat{\rho} e^{\hat{c}^\dagger h}e^{-h^*\hat{c}}|k\rangle
e^{h^*g} e^{-g^* h} |g\rangle\langle g| \, .
\end{eqnarray}
Comparing this with the form $\int dg^*dg \: \phi(g,g^*) |g\rangle\langle g|$ leads, directly, to an 
explicit form for the Grassmann function $\phi(g,g^*)$:
\begin{equation}
\label{Eq49}
\phi(g,g^*) =  \int dh^*dh \int dk^*dk \langle k| \hat{\rho} e^{\hat{c}^\dagger h}e^{-h^*\hat{c}}|k\rangle
e^{h^*g} e^{-g^* h} \, .
\end{equation}

The second possibility is to insert the identity given in Eq (\ref{Eq33a}):
\begin{eqnarray}
\label{Eq50}
\hat{\rho} &=&  \int dh^*dh \int dk^*dk \langle k| \hat{\rho} e^{\hat{c}^\dagger h}e^{-h^*\hat{c}}|k\rangle
e^{h^*\hat{c}}\int dg^*dg (2gg^* - 1)|g\rangle\langle -g| e^{-\hat{c}^\dagger h}  \nonumber \\
&=& \int dg^*dg \int dh^*dh \int dk^*dk \langle k| \hat{\rho} e^{\hat{c}^\dagger h}e^{-h^*\hat{c}}|k\rangle
e^{h^*g} e^{g^* h} (2gg^* - 1)|g\rangle\langle -g| \, .  \nonumber \\
& & 
\end{eqnarray}
If we compare this with the form $\int dg^*dg \: P(g,g^*)|g\rangle\langle -g|$ then we obtain an explicit form 
of the Grassmann function $P(g,g^*)$:
\begin{equation}
\label{Eq51}
P(g,g^*) = \int dh^*dh \int dk^*dk \langle k| \hat{\rho} e^{\hat{c}^\dagger h}e^{-h^*\hat{c}}|k\rangle
e^{h^*g} e^{g^* h}(2gg^* - 1)  \, ,
\end{equation}
which clearly differs from the function $\phi(g,g^*)$.  The two P functions are simply related to each other,
however:
\begin{eqnarray}
\label{Eq52}
P(g,g^*) &=& (2gg^*-1)\phi(g,-g^*)  \nonumber \\
\phi(g,g^*) &=& (2gg^*-1)P(g,-g^*) \, .
\end{eqnarray}
That this is satisfactory follows directly from the condition that $(-2gg^*-1)(2gg^*-1) = 1$.

\section{Properties of the two P representations}

We have two quantities, $\phi(g,g^*)$ and $P(g,g^*)$, that are the analogues for fermions of the bosonic 
quasiprobability distribution $P(\alpha,\alpha^*)$.  Like $P(\alpha,\alpha^*)$, we cannot expect either of them to be
a true probability distribution.  This is, perhaps, yet clearer for our fermionic functions as they
depend on Grassmann variables, which have no interpretation as physical properties of the 
fermions.  Despite this, we can treat both $\phi(g,g^*)$ and $P(g,g^*)$ in a manner analogous to
probability distributions; in particular we can use them to evaluate occupation probabilities
and, in multimode situations, also correlation functions.  We examine here these features 
of $\phi(g,g^*)$ and $P(g,g^*)$.

\subsection{Normalisation?}

The integral of $P(\alpha,\alpha^*)$ over the whole of the complex plane gives unity:
\begin{equation}
\label{Eq53}
\int d^2\alpha \: P(\alpha,\alpha^*) = 1
\end{equation}
and although $P(\alpha,\alpha^*)$ can take negative values, this normalisation is indicative of its role
as a quasiprobability distribution  \cite{Radmore}.  We can use our explicit expressions for
$\phi(g,g^*)$ and $P(g,g^*)$ to determine whether either of these is normalised under Grassmann 
integration.  Let us start with $\phi(g,g^*)$:
\begin{eqnarray}
\label{Eq54}
\int dg^*dg \: \phi(g,g^*) &=& \int dh^*dh \int dk^*dk \langle k|\hat{\rho} e^{\hat{c}^\dagger h}e^{-h^*\hat{c}}|k\rangle
\int dg^*dg \: e^{h^*g}e^{-g^*h}  \nonumber \\
&=& \int dh^*dh \int dk^*dk \langle k|\hat{\rho} e^{\hat{c}^\dagger h}e^{-h^*\hat{c}}|k\rangle hh^* \, .
\end{eqnarray}
The final part, $hh^*$, behaves like a Dirac delta function in that it sets $h$ and $h^*$ to zero 
under the integral $\int dh^*dh$.  This leaves 
\begin{eqnarray}
\label{Eq55}
\int dg^*dg \: \phi(g,g^*) &=& \int dk^*dk \langle k|\hat{\rho}|k\rangle \nonumber \\
&=& \langle 0|\hat{\rho}|0\rangle - \langle 1|\hat{\rho}|1\rangle \, ,
\end{eqnarray}
which is not the trace of the density operator $\hat{\rho}$ and so is not generally equal to unity.  
Moreover, if the single particle probability exceeds the vacuum probability then this integral 
will be negative.  In order to fulfil the role of a quasiprobability distribution, we need
to introduce a weight function to include on the normalisation and moment integrals.  It is 
straightforward to show that this weight function is
\begin{equation}
\label{Eq55a}
w(g,g^*) = 2gg^* + 1 \, ,
\end{equation}
so that the product $w(g,g^*)\phi(g,g^*)$ is normalised:
\begin{equation}
\int dg^* dg w(g,g^*)\phi(g,g^*) = 1 \, .
\end{equation}
The necessity of introducing a weight function appears, also, in the bosonic case, where the identity 
is given by $\int \frac{d^2\alpha}{\pi} |\alpha\rangle\langle \alpha| = \hat{\rm I}$ but the normalisation condition
requires the weight function $\pi$: $\int \frac{d^2\alpha}{\pi} \: \pi\times P(\alpha,\alpha^*) = 1$.  For the 
fermions we have $\int dgdg^* |g\rangle\langle g| = \hat{\rm I}$ but $\int dgdg^* \: w(g,g^*)\phi(g,g^*) = 1$.

For $P(g,g*)$ we find
\begin{eqnarray}
\label{Eq56}
\int dg^*dg \: P(g,g^*) &=& \int dh^*dh \int dk^*dk \langle k| \hat{\rho} e^{\hat{c}^\dagger h}e^{-h^*\hat{c}}|k\rangle \nonumber \\
& &  \qquad \times\int dg^*dg \: e^{h^*g} e^{g^* h}(2gg^* - 1)  \nonumber \\
&=& \int dh^*dh \int dk^*dk \langle k| \hat{\rho} e^{\hat{c}^\dagger h}e^{-h^*\hat{c}}|k\rangle(2 + hh^*)  \nonumber \\
&=& \int dk^*dk \left(2\langle k|\hat{\rho}(-\hat{c}^\dagger \hat{c})|k\rangle + \langle k|\hat{\rho}|k\rangle \right) \nonumber \\
&=& \langle 0|\hat{\rho}|0\rangle + \langle 1|\hat{\rho}|1\rangle \, ,
\end{eqnarray}
which is the trace of the density operator and hence 
\begin{equation}
\label{Eq57}
\int dg^*dg \: P(g,g^*) = 1 \, .
\end{equation}
The function $P(g,g^*)$ has the normalisation attribute of a quasiprobability distribution, while $\phi(g,g^*)$
requires the addition of the weight function $w(g,g^*)$ to produce a normalised Grassmann function.

\subsection{Evaluating moments}
\label{mom}

For our single mode problem there are only two possible distinct moments.  These are the zeroth moment, $\langle\hat{\rm I}\rangle$,
and the first moment, $\langle\hat{c}^\dagger\hat{c}\rangle$, of the particle number operator.  Superpositions of zero and one 
fermion are not allowed and so $\langle\hat{c}^\dagger\rangle$ and $\langle\hat{c}\rangle$ are always zero.  We have two coherent-state
representations of the density operator for our single-mode state, and we can use either of these to evaluate moments.

We consider, first, the zeroth moment $\langle\hat{\rm I}\rangle$, which must equal unity.  For the representation in terms of
$\phi(g,g^*)$ we have
\begin{eqnarray}
\label{Eq58}
\langle\hat{\rm I}\rangle &=& {\rm Tr}\int dg^*dg \: \phi(g,g^*) |g\rangle\langle g| \nonumber \\
&=& \int dg^*dg \: \phi(g,g^*)(\langle 0|g\rangle\langle g|0\rangle + \langle 1|g\rangle\langle g|1\rangle) \nonumber \\
&=& \int dg^*dg \: \phi(g,g^*)(2gg^* +1) \, .
\end{eqnarray}
Note that the weight function, $w(g,g^*) = 2gg^* + 1$ appears naturally in this expression.
Here we have used the fact that $\phi(g,g^*)$ must be an even Grassmann function and so be formed only of products of 
even numbers of Grassmann variables.  Were this not the case, then we would have non-zero expectation values of
$\hat{c}$ and/or $\hat{c}^\dagger$.  For our Grassmann integral to equal unity (as it must) $\phi(g,g^*)$ must have the
highly restricted form
\begin{equation}
\label{Eq59}
\phi(g) = u +(1 - 2u)gg^* \, ,
\end{equation}
where $u$ is a constant.  For the representation in terms of $P(g,g^*)$ we have 
\begin{eqnarray}
\label{Eq60}
\langle\hat{\rm I}\rangle &=& \int dg^*dg \: P(g,g^*)(\langle 0|g\rangle\langle -g|0\rangle + \langle 1|g\rangle\langle -g|1\rangle)
\nonumber \\
&=& \int dg^*dg \: P(g,g^*) \, ,
\end{eqnarray}
which is the normalisation condition derived above.  As with $\phi(g,g^*)$, this condition restricts $P(g,g^*)$ to a simple form
\begin{equation}
\label{Eq61}
P(g,g^*) = v + gg^* \, ,
\end{equation}
where $v$ is a constant.  The simple forms of $\phi(g)$ and $P(g)$ arise because the allowed states of a single
fermionic mode are limited to statistical mixtures of the vacuum and one particle states: $\hat{\rho} = 
(1-p)|0\rangle\langle 0| + p|1\rangle\langle 1|$.  These forms follow also directly from Eqs (\ref{Eq49}) and 
(\ref{Eq51}) for a density operator given by this statistical mixture.  We see that both $P(g,g^*)$ and $\phi(g,g^*)$
are even Grassmann functions as a consequence of this super-selection rule.

The first order moment $\langle\hat{c}^\dagger\hat{c}\rangle$ is the probability, $p$, that a single fermion is present in
the mode.  For $\phi(g,g^*)$ we find 
\begin{eqnarray}
\label{Eq62}
\langle\hat{c}^\dagger\hat{c}\rangle &=& {\rm Tr}\int dg^*dg \: w(g,g^*)\phi(g,g^*)|g\rangle\langle g| \hat{c}^\dagger\hat{c} \nonumber \\
&=& {\rm Tr}\int dg^*dg \: w(g,g^*)\phi(g,g^*)\hat{c}|g\rangle\langle g| \hat{c}^\dagger \nonumber \\
&=& \int dg^*dg \: (2gg^* + 1)\phi(g,g^*)gg^* \nonumber \\
&=& \int dg^*dg \: \phi(g,g^*)gg^* \, .
\end{eqnarray}
For this to equal the single fermion probability we require $u = p$ so that $\phi(g,g^*) = p + (2p-1)gg^*$.  If we use the 
representation in terms of $P(g,g^*)$ then we find 
\begin{eqnarray}
\label{Eq63}
\langle\hat{c}^\dagger\hat{c}\rangle &=& {\rm Tr}\int dg^*dg \: P(g)|g\rangle\langle -g| \hat{c}^\dagger\hat{c} \nonumber \\
&=& {\rm Tr}\int dg^*dg \: P(g,g^*)\hat{c}|g\rangle\langle -g| \hat{c}^\dagger \nonumber \\
&=& \int dg^*dg \: P(g,g^*)g^*g \, .
\end{eqnarray}
For this expression to be the single fermion occupation probability we require $v = -p$ so that $P(g,g^*) = -p + gg^*$.
We note that, in this single mode case, the forms of the moments for our two suitably weighted
quasiprobability distributions, $w(g,g^*)\phi(g,g^*)$ and $P(g,g^*)$, differ 
only by the ordering of the Grassmann variables $g$ and $g^*$.  For the bosonic case, $\alpha$ is a c-number
and so there is no distinction between $\alpha^*\alpha$ and $\alpha\alpha^*$.

\subsection{Multimode multiparticle states}

The Pauli exclusion principle means that single-mode states are limited to the presence or the absence of a single
particle.  Most physically interesting states are therefore multimodal.  Cold atom states, for example, often span a
very large number of modes.  To describe such states we require multimode generalisations of the expressions derived
above.

The two important completeness relations, Eqs (\ref{Eq30}) and (\ref{Eq33a}), become
\begin{equation}
\label{Eq64}
\int d^2{\bf g} \: |{\bf g}\rangle\langle {\bf g}| = \hat{\rm I}
\end{equation}
and
\begin{equation}
\label{Eq65}
\int d^2{\bf g} \: f({\bf g},{\bf g}^*)|{\bf g}\rangle\langle -{\bf g}| = \hat{\rm I} \, ,
\end{equation}
where $f({\bf g},{\bf g}^*)$ is
\begin{equation}
\label{Eq66}
f({\bf g},{\bf g}^*) = \prod_i (2g_ig^*_i - 1) \, ,
\end{equation}
which is clearly an even Grassmann function, and $d^2{\bf g} \equiv \prod_i dg^*_idg_i$.  The multimode coherent states are
generated from the vacuum (or zero particle) state by the displacement operator $\hat{D}({\bf g},{\bf g}^*)$:
\begin{equation}
\label{Eq67}
|{\bf g},{\bf g}^*\rangle = \hat{D}({\bf g},{\bf g}^*)|{\bf 0}\rangle \, ,
\end{equation}
where 
\begin{equation}
\label{Eq68}
\hat{D}({\bf g},{\bf g}^*) = \exp\left(\hat{\bf c}^\dagger\cdot{\bf g} - {\bf g}^*\cdot\hat{\bf c}\right) \, .
\end{equation}
Here the vector quantities ${\bf g}$ and $\hat{\bf c}$ denote $\{g_1, g_2, \cdots, g_i, \cdots \}$ and
$\{\hat{c}_1, \hat{c}_2, \cdots , \hat{c}_i, \cdots\}$ respectively.  To condense the notation, $|{\bf g},{\bf g}^*\rangle$ and $\hat{D}({\bf g},{\bf g}^*)$
will be replaced by $|{\bf g}\rangle$ and $\hat{D}({\bf g})$.  It follows that our multimode coherent states and 
displacement operators are
\begin{eqnarray}
\label{Eq69}
|{\bf g}\rangle &=& \prod_i |g_i\rangle \nonumber \\
\hat{D}({\bf g}) &=& \prod_i \hat{D}(g_i) \, .
\end{eqnarray}
In the multimode case the projector expansion, Eq(\ref{Eq35}), becomes
\begin{eqnarray}
\label{Eq70}
|{\bf g}_1\rangle\langle {\bf g}_2| &=& \int d^2{\bf h} \int d^2{\bf k} \: \langle{\bf g}_2|\hat{D}({\bf h})|{\bf k}\rangle\langle {\bf k}|{\bf g}_1\rangle \hat{D}(-{\bf h})
\nonumber \\
&=& \int d^2{\bf h} \int d^2{\bf k} \: \langle {\bf k}|{\bf g}_1\rangle\langle{\bf g}_2|\hat{D}({\bf h})|{\bf k}\rangle \hat{D}(-{\bf h}) \, ,
\end{eqnarray}
so that our single-mode operator identity, Eq (\ref{Eq36}), becomes
\begin{equation}
\label{Eq71} 
\hat{F} = \int d^2{\bf h} \int d^2{\bf k} \: \langle{\bf k}|\hat{F}\hat{D}({\bf h}) |{\bf k}\rangle \hat{D}(-{\bf h}) \, .
\end{equation}

If we follow our single-mode analysis and insert the multimode identity, Eq (\ref{Eq64}), into the antinormally ordered form of 
$\hat{D}(-{\bf h})$ we find an explicit expression for the density operator in terms of the multimode quasiprobability distribution
$\phi({\bf g},{\bf g}^*)$:
\begin{equation}
\label{Eq72}
\hat{\rho} = \int d^2{\bf g} \: \phi({\bf g},{\bf g}^*) |{\bf g}\rangle\langle {\bf g}| \, ,
\end{equation}
where $\phi({\bf g},{\bf g}^*)$ is given by 
\begin{equation}
\label{Eq73}
\phi({\bf g},{\bf g}^*) \int d^2{\bf h}\int d^2{\bf k} \: \langle {\bf k}|\hat{\rho} e^{\hat{\bf c}^\dagger\cdot{\bf h}}e^{-{\bf h}^*\cdot\hat{\bf c}}|{\bf k}\rangle
e^{{\bf h}^*\cdot{\bf g}}e^{-{\bf g}^*\cdot{\bf h}} \, .
\end{equation}
We note that $\phi({\bf g},{\bf g}^*)$ is necessarily an even Grassmann function owing to the super-selection rule that the density operator
$\hat{\rho}$ has non-zero elements only between Fock states with either just an even or just an odd number of particles 
\footnote{The range of possible matrix elements depends very much on the physical system being modelled.  This is 
particularly true if we consider states of more than one type of fermion.  For cold
atoms, the number of atoms is a conserved quantity and so the only non-vanishing matrix elements will be those
between states with the same number of atoms.  In relativistic quantum theory, an intense electric field can 
create an electron-positron pair and thereby increase the number of fermions present by two.  Similarly, in
models of solids, an external electromagnetic field may act to create an electron-hole pair, again increasing the 
number of effective particles present by two.  If this is a coherent process, then the state can evolve to one in 
which there is a superposition of the vacuum state and a two particle state.}.

If we employ our second expression for the multimode identity operator, given in Eq (\ref{Eq65}), we obtain an expression
for the density operator in terms of our alternative multimode quasiprobability distribution:
\begin{equation}
\label{Eq74}
\hat{\rho} = \int d^2{\bf g} \: P({\bf g},{\bf g}^*) |{\bf g}\rangle\langle -{\bf g}| \, ,
\end{equation}
where $P({\bf g},{\bf g}^*)$ has the form
\begin{equation}
\label{Eq75}
P({\bf g},{\bf g}^*) = \int d^2{\bf h}\int d^2{\bf k} \: \langle {\bf k}|\hat{\rho} e^{\hat{\bf c}^\dagger\cdot{\bf h}}e^{-{\bf h}^*\cdot\hat{\bf c}}|{\bf k}\rangle
e^{{\bf h}^*\cdot{\bf g}}e^{{\bf g}^*\cdot{\bf h}} f({\bf g},{\bf g}^*) \, .
\end{equation}
We note that the super-selection rule ensures that $P({\bf g},{\bf g}^*)$ is also an even Grassmann function.
It follows that our two multimode quasiprobability distributions are related by
\begin{eqnarray}
\label{Eq76}
\nonumber
P({\bf g},{\bf g}^*) &=& f({\bf g},{\bf g}^*)\phi({\bf g},-{\bf g}^*) \nonumber \\
\phi({\bf g},{\bf g}^*) &=& f({\bf g},{\bf g}^*)P({\bf g},-{\bf g}^*) \, ,
\end{eqnarray}
the consistency of which follows from the fact that $f({\bf g},{\bf g}^*)f({\bf g},-{\bf g}^*) = 1$.

As with the single-mode case, the function $P({\bf g},{\bf g}^*)$ is normalised but $\phi({\bf g},{\bf g}^*)$, in general, is not:
\begin{eqnarray}
\label{Eq77}
\int d^2{\bf g} \: P({\bf g},{\bf g}^*) &=& 1 \nonumber \\
\int d^2{\bf g} \: \phi({\bf g},{\bf g}^*) &=& \int d^2{\bf k} \: \langle {\bf k}|\hat{\rho}|{\bf k}\rangle \, .
\end{eqnarray}
If we introduce the multimode weight factor, $w({\bf g},{\bf g}^*) = \prod_i(2g_ig_i^* + 1)$, then we obtain a 
normalisation condition for $\phi({\bf g},{\bf g}^*)$:
\begin{equation}
\label{Eq77a}
\int d^2{\bf g} \: w({\bf g},{\bf g}^*)\phi({\bf g},{\bf g}^*) = 1 \, .
\end{equation}
We can use both quasiprobability distributions to evaluate expectation values and correlations functions.

Our fermionic P representations allow us to calculate normally ordered expectation values.  To determine the forms of these,
it is convenient to start with the following operator identity
\begin{equation}
\label{Eq78}
\hat{c}_{m_q} \cdots \hat{c}_{m_1}\hat{\rho}\hat{c}^\dagger_{l_1} \cdots \hat{c}^\dagger_{l_p} =
\int d^2{\bf g} \: (g_{m_q} \cdots g_{m_1})P({\bf g},{\bf g}^*)[(-g^*_{l_1})\cdots (-g^*_{l_p})]|{\bf g}\rangle\langle -{\bf g}| \, ,
\end{equation}
which follows from the eigenvalue property of the coherent states together with the fact that $P({\bf g},{\bf g}^*)$
and $|{\bf g}\rangle\langle -{\bf g}|$ are both even Grassmann functions.  
The trace of this is the correlation function
\begin{equation}
\label{Eq78a}
{\rm Tr}\left(\hat{c}_{m_q} \cdots \hat{c}_{m_1}\hat{\rho}\hat{c}^\dagger_{l_1} \cdots \hat{c}^\dagger_{l_p} \right) =
{\rm Tr}\left(\hat{\rho}\hat{c}^\dagger_{l_1} \cdots \hat{c}^\dagger_{l_p} \hat{c}_{m_q} \cdots \hat{c}_{m_1}\right) \, ,
\end{equation}
where the equality holds by virtue of the cyclic property of the trace.  To determine the form of this trace,
it is helpful to consider first the trace of a general operator
\begin{equation}
\label{Eq78b}
\hat{F} = \int d^2{\bf g} F({\bf g},{\bf g}^*)|{\bf g}\rangle\langle -{\bf g}| \, ,
\end{equation}
where $F({\bf g},{\bf g}^*)$ is a general Grassmann function, which we may split into the sum of an even
and an odd Grassmann function.  These contain products of only even or odd numbers of Grassmann 
variables respectively:
\begin{equation}
\label{Eq78c}
F({\bf g},{\bf g}^*) = F({\bf g},{\bf g}^*)_e + F({\bf g},{\bf g}^*)_o \, .
\end{equation}
As we have seen, the  coherent states are even and so commute with Grassmann variables and it follows that
\begin{equation}
\label{Eq78d}
\hat{F} = \int d^2{\bf g}\:  |{\bf g}\rangle F({\bf g},{\bf g}^*)\langle -{\bf g}| \, .
\end{equation}
To proceed, it is simplest and clearest to specialise to a single mode, before generalising to the full multimode
problem.  For a single-mode operator, $\hat{F}_1$, we have
\begin{eqnarray}
\label{Eq78e}
{\rm Tr}\left(\hat{F}_1\right) &=& \int  d^2 g \left[\langle 0|g\rangle F_1(g,g^*)\langle -g|0\rangle 
+ \langle 1|g\rangle F_1(g,g^*)\langle -g|1\rangle \right]  \nonumber \\
&=& \int d^2 g \left[(1 + g g^*)F_1(g,g^*) - gF_1(g,g^*)g^*\right]  \nonumber \\
&=& \int d^2 g \: F_{1,e}(g,g^*)  \, ,
\end{eqnarray}
where we have used the fact that there are two Grassmann integrations and so the integral of any odd Grassmann function,
such as $(1 + g g^*)F_{1,o}(g,g^*)$ and $gF_{1,o}(g,g^*)g^*$, must be zero.

For the multimode case, we can write the trace of $\hat{F}$, in Eq (\ref{Eq78d}), as
\begin{equation}
\label{Eq78fa}
{\rm Tr}\left(\hat{F}\right) = \sum_{\nu_1,\cdots ,\nu_n} \int d^2{\bf g} \: \langle \nu_1,\cdots ,\nu_n|{\bf g}\rangle
F({\bf g},{\bf g}^*) \langle -{\bf g}|\nu_1,\cdots ,\nu_n\rangle \, ,
\end{equation}
where $|\nu_i\rangle = |0_i\rangle , \: |1_i\rangle$ are the fermion number states for mode $i$.  
To proceed further we write $F({\bf g},{\bf g}^*)$ as the sum of even and odd Grassmann functions (as in Eq (\ref{Eq78c}))
so that the right hand side of Eq (\ref{Eq78fa}) is the sum of two terms.  For the term involving $F_e({\bf g},{\bf g}^*)$ we
can commute $\langle \nu_1,\cdots,\nu_n|{\bf g}\rangle$ with $F_e({\bf g},{\bf g}^*)$ so that the first term becomes 
$\int d^2{\bf g} F_e({\bf g},{\bf g}^*)\sum_{\nu_1, \cdots, \nu_n}\langle \nu_1, \cdots, \nu_n|{\bf g}\rangle \langle -{\bf g}|\nu_1, \cdots, \nu_n\rangle$.
For the term involving $F_o({\bf g},{\bf g}^*)$
no simple commutation is possible and this contribution is just $\sum_{\nu_1,\cdots ,\nu_n} \int d^2{\bf g} \: \langle \nu_1,\cdots ,\nu_n|{\bf g}\rangle
F_o({\bf g},{\bf g}^*) \langle -{\bf g}|\nu_1,\cdots ,\nu_n\rangle$.  However, for each term in the sum over fermion occupancies
$\nu_1, \cdots, \nu_n$, the overlap $\langle \nu_1,\cdots,\nu_n|{\bf g}\rangle$ is the product of factors 
$\delta_{\nu_i,0}(1 + g_ig^*_i/2) + \delta_{\nu_i,1}g_i$ for each mode $i$, whilst $\langle-{\bf g}|\nu_1, \cdots, \nu_n\rangle$  
is the product of factors $\delta_{\nu_i,0}(1 + g_ig^*_i/2) - \delta_{\nu_i,1}g_i^*$ for each mode $i$.  Irrespective of whether $\nu_i$ 
is $0$ or $1$, for each term in the sum over fermion occupancies $\nu_1, \cdots, \nu_n$ and for each mode $i$, $F_o({\bf g},{\bf g}^*)$ 
will be multiplied by two factors, which are both odd (when $\nu_i = 1$) or both even (when $\nu_i = 0$).  The resulting Grassmann
functions for every term in the sum over fermion occupancies $\nu_1, \cdots, \nu_n$ will therefore still be odd.  As the 
integral of an odd Grassmann function is zero, it follows that the entire second term involving $F_o({\bf g},{\bf g}^*)$ will be zero.
This means that we can write the trace in the form
\begin{eqnarray}
\label{Eq78fb}
{\rm Tr}\left(\hat{F}\right) &=&  \int d^2{\bf g} \: 
F_e({\bf g},{\bf g}^*) \sum_{\nu_1,\cdots ,\nu_n} 
\langle \nu_1,\cdots ,\nu_n|{\bf g}\rangle \langle -{\bf g}|\nu_1,\cdots ,\nu_n\rangle \nonumber \\
&=&  \int d^2{\bf g} \: F_e({\bf g},{\bf g}^*)  
\prod_i \left(\langle 0_i|g_i\rangle\langle -g_i|0_i\rangle +   \langle 1_i|g_i\rangle\langle -g_i|1_i\rangle\right)
\nonumber \\
&=& \int d^2{\bf g} \: F_e({\bf g},{\bf g}^*)  \, .
\end{eqnarray}
For the present case (see Eqs (\ref{Eq78}) and (\ref{Eq78b})) $F({\bf g},{\bf g}^*)$ is given by 
$(g_{m_q} \cdots g_{m_1})P({\bf g},{\bf g}^*)$ $((-g_{l_1}) \cdots (-g_{l_p}))$.  The physical constraint of the super-selection 
rule on fermion number requires $P({\bf g},{\bf g}^*)$ to be an even Grassmann function.  It follows, therefore, that 
$F({\bf g},{\bf g}^*)$ is an even function if $p$ and $q$ differ by an even number and $F({\bf g},{\bf g}^*)$ is an odd
function if they differ by an odd number.  Hence $F_e({\bf g},{\bf g}^*) = F({\bf g},{\bf g}^*) = 
(g_{m_q} \cdots g_{m_1})P({\bf g},{\bf g}^*)((-g_{l_1} \cdots (-g_{l_p}))$ if $p-q$ is even and $F_e({\bf g},{\bf g}^*) = 0$
if $p-q$ is odd.  Returning to our correlation function we therefore find
\begin{eqnarray}
{\rm Tr}\left(\hat{\rho}\hat{c}^\dagger_{l_1} \cdots \hat{c}^\dagger_{l_p} \hat{c}_{m_q} \cdots \hat{c}_{m_1}\right) &=&
\int d^2{\bf g} \left[(g_{m_q} \cdots g_{m_1})P({\bf g},{\bf g}^*)(-g^*_{l_1})\cdots (-g^*_{l_p})\right]_e  \nonumber \\
& &   \qquad \qquad\qquad \qquad (p-q) \: {\rm even}
\nonumber \\
&=& 0 \qquad \qquad \qquad \qquad (p-q) \: {\rm odd} \, .
\end{eqnarray}
Thus the correlation function is zero unless $p$ and $q$ differ by an even number.  This corresponds to the requirement 
that the number of annihilation and creation operators in any non-zero correlation function must differ by an even number,
which includes zero of course.  For the result where $p-q$ is even, we can move all the $g_{m_i}$ to the right of the 
$g_{l_i}^*$ terms and $P({\bf g},{\bf g}^*)$ and remove the minus signs.  The factors $(-1)^{pq}$ and $(-1)^p$ involved
cancel when $(p-q)$ is even.  It follows that the normally ordered correlation function is
\begin{equation}
\label{Eq80}
{\rm Tr}\left(\hat{\rho}\hat{c}^\dagger_{l_1} \cdots \hat{c}^\dagger_{l_p}\hat{c}_{m_q} \cdots \hat{c}_{m_1}\right)
= \int d^2{\bf g} \: P({\bf g},{\bf g}^*)(g^*_{l_1}\cdots g^*_{l_p}) (g_{m_q} \cdots g_{m_1}) \, .
\end{equation}
This formula is actually true for $(p-q)$ both even and odd, as in the latter case both sides are zero.

We can proceed in a similar fashion with our second P representation, noting that $\phi({\bf g},{\bf g}^*)$ is
also an even function and that we only need to move it to the left of the $g_{m_i}$ terms.  We 
need, also, to include the multimode weight function $w({\bf g},{\bf g}^*)$.  The weight function is given by
$\prod_i (\langle 0_i|g_i\rangle\langle +g_i|0_i\rangle + \langle 1_i|g_i\rangle\langle +g_i|1_i\rangle)$
which replaces the product in the second line of Eq (\ref{Eq78fb}) for the $\phi({\bf g},{\bf g}^*)$ case.  Here there is no $(-1)^p$ 
factor.  We find  
\begin{eqnarray}
\label{Eq81}
{\rm Tr}\left(\hat{\rho}\hat{c}^\dagger_{l_1} \cdots \hat{c}^\dagger_{l_p}\hat{c}_{m_q} \cdots \hat{c}_{m_1}\right)
&=& \int d^2{\bf g} \: (g_{m_q} \cdots g_{m_1})w({\bf g},{\bf g}^*)\phi({\bf g},{\bf g}^*)(g^*_{l_1} \cdots g^*_{l_p}) \nonumber \\
&=& \int d^2{\bf g} \: w({\bf g},{\bf g}^*)\phi({\bf g},{\bf g}^*)(g_{m_q} \cdots g_{m_1})(g^*_{l_1} \cdots g^*_{l_p}) \, , \nonumber \\
& & 
\end{eqnarray}
the formula being true for $(p-q)$ both even and odd.  For both representations the Grassmann integrals on the 
right-hand sides of Eqs (\ref{Eq80}) and (\ref{Eq81}) will be zero if $p$ and $q$ differ by an odd number, as 
the Grassmann integral of an odd function must be zero.  This corresponds to the correlation function itself being zero in this case.
For our simple analysis in which we consider only one type of fermion, the density operator for states that comply with 
the super-selection rule has only non-zero matrix elements between Fock states with same total fermion number,
and hence is zero unless $p=q$.  Note that if $p=0=q$ then Eqs (\ref{Eq80}) and (\ref{Eq81}) reproduce the 
normalisation conditions for $P({\bf g},{\bf g}^*)$ and $\phi({\bf g},{\bf g}^*)$ in Eqs (\ref{Eq77}) and (\ref{Eq77a}).
The differences between the two forms of the correlation functions and, in particular, the presence or absence of a
weight function is due, ultimately, to the two different resolutions of the identity, Eqs. (\ref{Eq30}) and (\ref{Eq33a})


The difference between our two forms for the correlation functions, Eqs (\ref{Eq80}) and (\ref{Eq81}), finally makes 
clear the reason why there are two fermionic P representations but only a single bosonic one.  In the bosonic case,
the quantities $\alpha_i$ and $\alpha^*_j$ are c-numbers and so commute.  In the fermionic case, however, the quantities 
$g_i$ and $g^*_j$ are Grassmann variables, which means that the order of these matters.  Correlation functions 
calculated using $P({\bf g},{\bf g}^*)$ evaluate these with the conjugate Grassmann variables, $g^*_i$ to the left of the 
non-conjugated ones, $g_i$, while for $w({\bf g},{\bf g}^*)\phi({\bf g},{\bf g}^*)$ it is the other way around.  The results
in Eqs (\ref{Eq80}) and (\ref{Eq81}) are seen to be the same using the relationship, Eq (\ref{Eq76}), between 
$P({\bf g},{\bf g}^*)$ and $\phi({\bf g},{\bf g}^*)$.  They are also consistent with the single mode results in Sect
\ref{mom}.


\section{Conclusion}

We have presented a derivation of the P representation that is valid both for bosons and for fermions.  A key feature 
of this derivation is the use of an expression for the identity operator, written as an integral over coherent states.
This leads to expressions for the density operator of the form
\begin{equation}
\label{Eq82}
\hat{\rho} = \int d^2\alpha \: P(\alpha,\alpha^*)|\alpha\rangle\langle \alpha| 
\end{equation}
for bosons and two forms,
\begin{eqnarray}
\label{Eq83}
\hat{\rho} &=& \int d^2 g \: P(g, g^*)|g\rangle \langle -g| \nonumber \\
&=& \int d^2g \: \phi(g,g^*)|g\rangle \langle g| \, ,
\end{eqnarray}
for fermions.  The existence of two P representations for fermions derives, simply, from the anticommuting property
of the Grassmann variables.  The quasiprobability function $P(g,g^*)$ should be used in conjunction with
Grassmann variables ordered so that $g^*$ occurs to the left of the $g$.  The function $\phi(g,g^*)$,
however, requires a weight function, $w(g,g^*)$, and the Grassmann variables in the other order.  
We have seen that this difference holds, also, for
multimode and multi-particle fermionic states.  The correlation functions given in Eqs (\ref{Eq80}) and (\ref{Eq81})
demonstrate the generality of this conclusion.  In dynamical problems, it may be more natural to employ
$P(g,g^*)$ than $\phi(g,g^*)$ because although both have simple correspondence rules, it is harder to determine 
correlation functions in the $\phi({\bf g},{\bf g}^*)$ case due to the weight function $w({\bf g},{\bf g}^*)$.  It is certainly
possible to use either though.  It is also
possible to introduce a generalised P representation with two distinct Grassmann variables, $g$ and $g^+$ and
the P function $P(g,g^+)$ \cite{DJB}.  Such an approach has been used to determine the coherence properties 
between two Cooper pair states of a two-mode system \cite{Kidwani}.  Phase space theories for fermions involving
c-number phase space variables and Gaussian state projectors have also been introduced \cite{Corney}.  Here the
c-number elements of the covariance matrix involved in the Gaussian state act as phase space variables.  Finally, it has 
been shown that there exists an underlying c-number interpretation of the Grassmann phase space theory \cite{Polyakov}.

The theory presented here may readily be generalised to treat 
systems of several types of bosons and fermions.  A second type of boson would be represented by a new set of
commuting complex variables, {\boldmath{$\beta$}}, and a second type of fermion by a set of anti-commuting Grassmann
variables ${\bf h}$.  Each type would be associated with its own set of modes.


\begin{acknowledgments}

\noindent It is a pleasure to dedicate this work to Igor Jex on the occasion of his sixtieth birthday.  
This work was supported by the Royal Society via a Research Professorship (RP150122).

\end{acknowledgments}




\begin{thebibliography}{99}
\bibitem{footnote}
It is common practice to denote explicitly only the dependence on $\alpha$ by writing the quasiprobability distribution
as $P(\alpha)$.  We shall find it useful in our presentation of the properties of the corresponding fermionic functions
to have both variables present and so include both $\alpha$ and $\alpha^*$ in the argument of our bosonic quasiprobability
and both $g$ and $g^*$ in the corresponding fermionic functions.
\bibitem{footnote1}
It is interesting to note that the 
doubling of options for the phase space representation of fermionic operators 
applies generally.  One simple example is the parity operator, for which we have 
\begin{eqnarray}
(-\hat{\rm I})^{\hat{a}^\dagger\hat{a}} &=& \sum_{n=0}^\infty (-1)^n|n\rangle\langle n| \nonumber \\
&=& \int \frac{d^2\alpha}{\pi}|\alpha\rangle\langle -\alpha| \nonumber
\end{eqnarray}
for bosons.  For fermions we have the two representations:
\begin{eqnarray}
(-\hat{\rm I})^{\hat{c}^\dagger\hat{c}} &=& |0\rangle\langle 0| - |1\rangle \langle 1| \nonumber \\
&=& \int dg^*dg |g\rangle\langle -g| \nonumber \\
&=& \int dg^*dg(2gg^* - 1)|g\rangle\langle g| \, . \nonumber
\end{eqnarray}
\end{thebibliography}
\end{document}